# Inhomogeneous magnetic behavior of $Pr_{0.7}Ca_{0.3}CoO_3$ and $Nd_{0.7}Ca_{0.3}CoO_3$

**Asish K. Kundu[a], E.V. Sampathkumaran[b], K.V. Gopalakrishnan[b], C.N.R. Rao[a,*]**

[a] *Chemistry and Physics of Materials Unit, Jawaharlal Nehru Centre for Advanced Scientific Research, Jakkur PO, Bangalore - 560064, India.*
[b] *Department of Condensed Matter Physics & Materials Science, Tata Institute of Fundamental Research, Mumbai - 400 005, India.*

**Abstract**

Unlike $La_{0.7}Ca_{0.3}CoO_{2.97}$, $Pr_{0.7}Ca_{0.3}CoO_3$ and $Nd_{0.7}Ca_{0.3}CoO_{2.95}$ do not show distinct ferromagnetic transitions, but instead they exhibit very low magnetic moments down to 50 K. A detail study of magnetic properties of $Pr_{0.7}Ca_{0.3}CoO_3$ and $Nd_{0.7}Ca_{0.3}CoO_{2.95}$ shows that the materials are inhomogeneous, exhibiting properties similar to those of frustrated magnetic systems. In both these cobaltates, small ferromagnetic clusters seem to be present in an antiferromagnetic host.

[*]Corresponding author. Tel.: +91-80-23623075; fax: +91-80-23622760

*E-mail address:* cnrrao@jncasr.ac.in (C.N.R. Rao).

# 1. Introduction

Investigations of rare earth manganates of the general formula $Ln_{1-x}A_xMnO_3$ (Ln = rare earth, A = alkaline earth) have revealed remarkable aspects of these materials which include colossal magnetoresistance (CMR), charge ordering (CO), orbital ordering as well as electronic phase separation [1-4]. Properties of these materials are crucially controlled by the average radius of the A-site cation, $<r_A>$ [3-7]. Thus, while $La_{0.7}Ca_{0.3}MnO_3$ ($<r_A>$ = 1.205 Å) shows insulator to metal transition and ferromagnetism, with metallicity associated with ferromagnetism at low temperatures [3, 4, 8], $Pr_{0.7}Ca_{0.3}MnO_3$ with a smaller $<r_A>$ (1.179 Å) shows no ferromagnetism or insulator to metal transition. Instead, the latter manganate exhibits charge ordering, orbital ordering and electronic phase separation [3, 4, 9]. We were interested in exploring whether the analogous cobaltates of general formula $Ln_{1-x}A_xCoO_3$ (Ln = rare earth, A = alkaline earth) exhibit similar features. In this cobaltate system, the x = 0.3 composition is ferromagnetic and metallic when Ln = La and A = Sr or Ca [10-15] and also when Ln = Pr and A = Sr [15-17]. A cluster-glass behavior has also been suggested at low temperatures in these materials [11], but this has not been entirely established [14]. We considered it important to investigate the properties of $Pr_{0.7}Ca_{0.3}CoO_3$ and $Nd_{0.7}Ca_{0.3}CoO_3$, with smaller $<r_A>$ values of 1.179 and 1.168 Å respectively, to examine how their properties vary from those of $La_{0.7}Ca_{0.3}CoO_3$ ($<r_A>$ = 1.354 Å).

## 2. Experimental Procedure

Polycrystalline samples of $Ln_{0.7}Ca_{0.3}CoO_{3-\delta}$ (Ln = La, Pr, Nd) were prepared by the conventional ceramic method. Stoichiometric mixtures of the respective rare earth oxides, alkaline earth carbonates and $Co_3O_4$ were weighed in desired proportions and milled for few hours with propanol. After the mixed powders were dried, they were calcined in air at 950 °C followed by heating at 1000 and 1100 °C for 12h each in air. The powders thus obtained were pelletized and the pellets sintered at 1200 °C for 12h in air. To improve the oxygen stoichiometry the samples were annealed in an oxygen atmosphere at a lower temperature ($\leq$ 900 °C). The oxygen stoichiometry was determined by iodometric titrations, the error in the determination being ± 0.02. The oxygen stoichiometry in the $Ln_{0.7}Ca_{0.3}CoO_{3-\delta}$ (Ln = La, Pr, Nd) thus obtained were 2.97, 3.00 and 2.95 respectively in the La, Pr and Nd derivatives.

The phase purity of the samples was established by recording the X-ray diffraction patterns in the 2θ range of 10°-80° with a Seiferts 3000 TT diffractometer, employing Cu-Kα radiation. The unit cell parameters of $Ln_{0.7}Ca_{0.3}CoO_{3-\delta}$ (Ln = La, Pr, Nd) are listed in Table 1 along with the weighted average radius $<r_A>$. The $<r_A>$, values were calculated using the Shannon radii for 12-coordination in the case of rhombohedral cobaltates, and for 9-coordination in the case of the orthorhombic ones. The uncertainties in the unit cell parameters are ± 0.004 Å. Magnetization measurements were made with a vibrating sample magnetometer (Lakeshore 7300) and with a SQUID magnetometer (Quantum Design). Electrical resistivity (ρ) measurements were carried out by the four-probe method with silver epoxy as the electrodes in the 300 – 20 K temperature ranges.

## 3. Results and discussion

Preliminary measurements (at 1 kOe) of the DC magnetic susceptibilities of $Ln_{0.7}Ca_{0.3}CoO_{3-\delta}$ with Ln = La, Pr, Nd showed that while $La_{0.7}Ca_{0.3}CoO_{2.97}$ clearly exhibits a ferromagnetic transition ($T_c$ ~175 K), $Pr_{0.7}Ca_{0.3}CoO_3$ and $Nd_{0.7}Ca_{0.3}CoO_{2.95}$ do not show distinct ferromagnetic transitions down to 50 K (Fig. 1a). There is a slight increase in the susceptibility of $Pr_{0.7}Ca_{0.3}CoO_3$ around 75 K, but this is not due to a genuine ferromagnetic transition. The magnetic behavior of a single crystal of $Pr_{0.7}Ca_{0.3}CoO_3$ is similar to that of the polycrystalline sample (see inset of Fig. 1a). On the basis of the $<r_A>$ values, the ferromagnetic $T_c$'s of $Pr_{0.7}Ca_{0.3}CoO_3$ and $Nd_{0.7}Ca_{0.3}CoO_3$ would be expected to be well above 100 K. Electrical resistivities of these cobaltates are also much higher (Fig. 1b). The large drop in the magnetic moment at low temperatures in the Pr and Nd derivatives is noteworthy. In order to understand the nature of these materials, we have investigated the magnetic properties of $Pr_{0.7}Ca_{0.3}CoO_3$ in detail, down to low temperatures.

In Fig. 2 we show the temperature variation of the DC magnetic susceptibility of $Pr_{0.7}Ca_{0.3}CoO_3$ in the zero-field-cooled (ZFC) and field-cooled (FC) states (H = 100 Oe). There is considerable divergence in the ZFC and FC behavior just as in magnetically frustrated systems [11]. The data show two broad transitions around 60 K and 30 K. Measurements carried out at 5 kOe, however, do not reveal the two peaks (Fig 3), suggesting that the intermediate temperature range M-H behavior of this material is rather complex at low fields. The data in Fig. 3 suggest that magnetic ordering sets in around 75 K with the susceptibility going through a broad maximum around 15 K. Inverse magnetic susceptibility data, shown in the inset of Fig 3, yield a Curie temperature ($\theta_p$) of - 30 K. The high temperature linear region of the inverse susceptibility data gives a magnetic moment of 4.7 $\mu_B$. The shape of $\chi - T$ plot below 75 K is rather complex, not typical of normal ferromanets. It appears as though there is a spread of magnetic transition temperatures due to local environmental effects.

In Fig. 4, we show the M – H behavior of $Pr_{0.7}Ca_{0.3}CoO_3$. The behavior is rather complex especially in the temperature range of 25 – 60 K. The plots remain nonlinear upto 120 kOe even at 5 K. The behavior is unlike of ferromagnets and is somewhat comparable to that of frustrated systems. Extrapolation of the M – H data in the high field region to zero field gives a saturation moment of around 0.4 $\mu_B$. The small value of the moment on cobalt in the apparently ferromagnetic state, compared with the value in the paramagnetic state, indicates itinerant ferromagnetism, which is possible because the material is conducting. From Fig. 5 we see that there is hysteresis at 5 K even at low fields, suggesting a ferromagnet-like behavior. The width of the hysteresis loop decreases markedly with increasing temperature. The above results reveal that ferromagnetic and antiferromagnetic interactions coexist at low temperatures, with the small conducting ferromagnetic domains or clusters giving rise to a small magnetic moment.

AC susceptibility measurements (Fig. 6) show that the low-temperature transition has a frequency dependence of about 1.3 K, as the frequency is increased from 1.3 to 1330 Hz. The 60 K peak, however, shows little shift (Fig. 6). The position of the low temperature peak in the AC susceptibility data at 1.3 Hz, for which the field of measurement is 1 Oe, occurs at 37.4 K, and shifts to lower temperatures at higher fields.

Thus, for H = 100 and 5000 Oe, the peak occurs at 31.5 and 12.7 K respectively. Because of the inhomogeneous nature, it is difficult to clearly assign one temperature for the bulk transition in this cobaltate, although the first transition clearly occurs around 60 K. While we have compared the inhomogeneous nature of $Pr_{0.7}Ca_{0.3}CoO_3$ at low temperatures to that of cluster or spin-glasses [11], isothermal remnant magnetization measurements in the 5 – 60 K range rule out that the material is actually a glass. Thus, isothermal remnant magnetization is time-independent and does not decay logarithmically or exponentially. We, therefore, conclude that the behavior of $Pr_{0.7}Ca_{0.3}CoO_3$ represents a special case of electronic phase separation.

We have carried out studies on polycrystalline $Nd_{0.7}Ca_{0.3}CoO_{2.95}$ as well. This sample also shows divergence in the ZFC and FC behavior at H = 100 Oe (Fig. 7), but the divergence is not marked as much as in $Pr_{0.7}Ca_{0.3}CoO_3$. The ZFC data seems to suggest two close transitions between 0 and 20 K. The DC susceptibility data at high fields (H = 5 kOe) shows one distinct transition around 20 K (Fig. 8). The inverse magnetic susceptibility data yield a $\theta_p$ value – 170 K. The M – H behavior of this cobaltate is also nonlinear just as $Pr_{0.7}Ca_{0.3}CoO_3$. The material shows narrow hysteresis below 5 K and below, at high fields (see inset Fig. 8).

The electronic phase separation and associated magnetic properties of $Pr_{0.7}Ca_{0.3}CoO_3$ and $Nd_{0.7}Ca_{0.3}CoO_{2.95}$ arise because of the small average size of the A-site cations. In these two cobaltates, the average radius (for orthorhombic structure) is less than 1.18 Å, which is the critical value only above which long-range ferromagnetism manifests itself [18]. It is known that increase in size disorder and decrease in size favor phase separation.

## 4. Conclusions

$Pr_{0.7}Ca_{0.3}CoO_3$ does not show a sharp ferromagnetic transition down to 50 K. There is large divergence between the DC magnetic susceptibility of the ZFC and FC sample. The magnetization is nonlinear with field. AC susceptibility data show evidence for a magnetic transition around 60 K and a frequency-dependent transition at low temperatures. Isothermal remnant magnetization measurements, however, reveal that the cobaltate is not a spin-glass. Properties of $Nd_{0.7}Ca_{0.3}CoO_{2.95}$ are not unlike those of the Pr analogue. These various features indicate that these cobaltates are magnetically inhomogeneous, with small ferromagnetic clusters or domains being present in an antiferromagnetic matrix.

**Acknowledgements**

The authors would like to thank BRNS (DAE), India for support of this research. A. K. K wants to thank University Grants Commission, India for fellowship award.

**Figure captions**

**Fig. 1** Temperature dependence of (a) the magnetic susceptibility, $\chi$, (H = 1000 Oe) and (b) the electrical resistivity, $\rho$, of $Ln_{0.7}Ca_{0.3}CoO_{3-\delta}$ (Ln = La, Pr or Nd). The inset in (a) shows the magnetic susceptibility, $\chi$, and inverse magnetic susceptibility, $\chi^{-1}$, of $Pr_{0.7}Ca_{0.3}CoO_3$ for single crystal and polycrystalline samples.

**Fig. 2** Temperature dependence of magnetic susceptibility, $\chi$, of $Pr_{0.7}Ca_{0.3}CoO_3$ (H = 100 Oe). Solid and dotted lines represent zero-field-cooled (ZFC) and field-cooled (FC) data respectively.

**Fig. 3** Temperature dependence of magnetic susceptibility, $\chi$, (H = 5000 Oe) of $Pr_{0.7}Ca_{0.3}CoO_3$. The inset shows the temperature dependence of inverse magnetic susceptibility, $\chi^{-1}$, (H = 5000 Oe).

**Fig. 4** The high field magnetization curve of $Pr_{0.7}Ca_{0.3}CoO_3$ at low temperatures.

**Fig. 5** Low field magnetic hysteresis of $Pr_{0.7}Ca_{0.3}CoO_3$ at low temperatures.

**Fig. 6** AC- magnetic susceptibility data of $Pr_{0.7}Ca_{0.3}CoO_3$ at 1 Oe.

**Fig. 7** Temperature dependence of the magnetic susceptibility, $\chi$, of $Nd_{0.7}Ca_{0.3}CoO_{2.95}$ (H = 100 Oe). Solid and dotted lines represent zero-field-cooled (ZFC) and field-cooled (FC) data respectively.

**Fig. 8** Temperature dependence of magnetic susceptibility, $\chi$, of $Nd_{0.7}Ca_{0.3}CoO_{2.95}$ (H = 5000 Oe). The inset shows low temperatures hysteresis.

**Table 1. Crystal Structure data of $Ln_{0.7}Ca_{0.3}CoO_{3-\delta}$ (Ln = La, Pr, Nd)**

| Composition | $\langle r_A \rangle$ (Å) | Space group | Lattice parameters (Å) | | | V (Å$^3$) |
| --- | --- | --- | --- | --- | --- | --- |
| | | | a | b | c | |
| $La_{0.7}Ca_{0.3}CoO_{2.97}$ | 1.354 | $R\bar{3}C$ | 5.3906 | - | - | 111.60 |
| $Pr_{0.7}Ca_{0.3}CoO_{3.00}$ | 1.179 | Pnma | 5.3577 | 7.5774 | 5.3436 | 216.94 |
| $Nd_{0.7}Ca_{0.3}CoO_{2.95}$ | 1.168 | Pnma | 5.3460 | 7.5638 | 5.3287 | 215.47 |

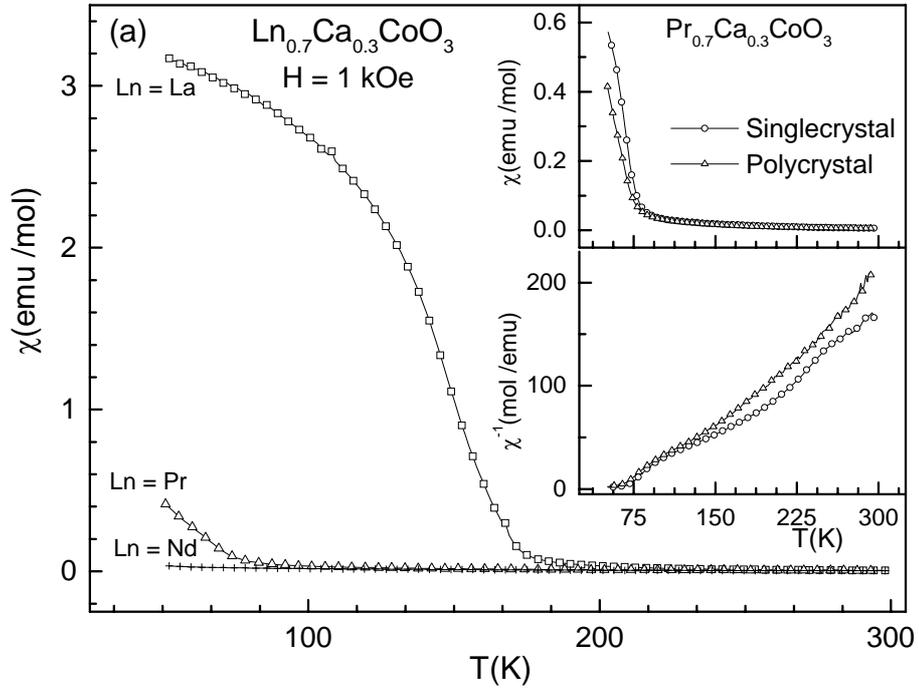

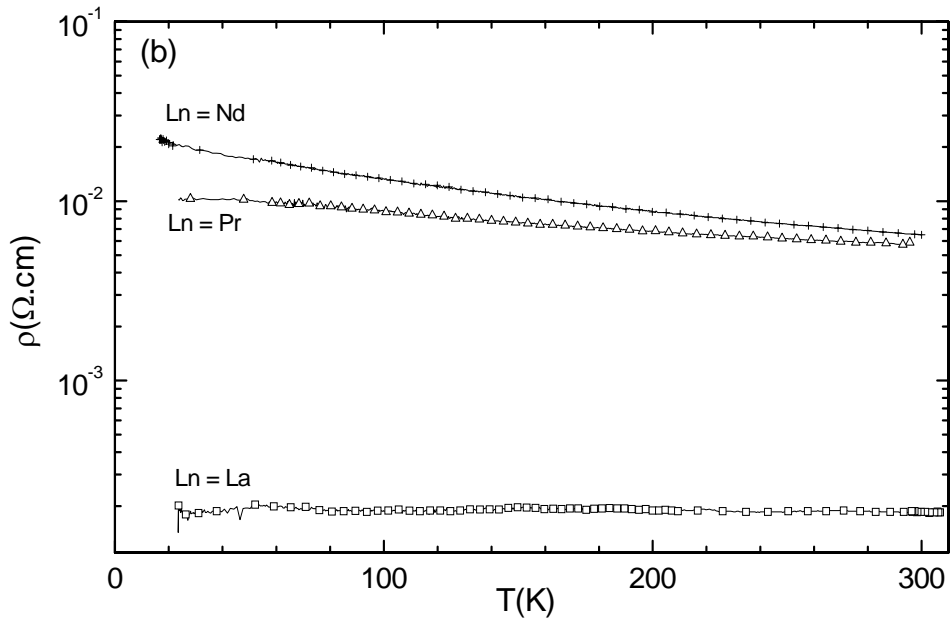

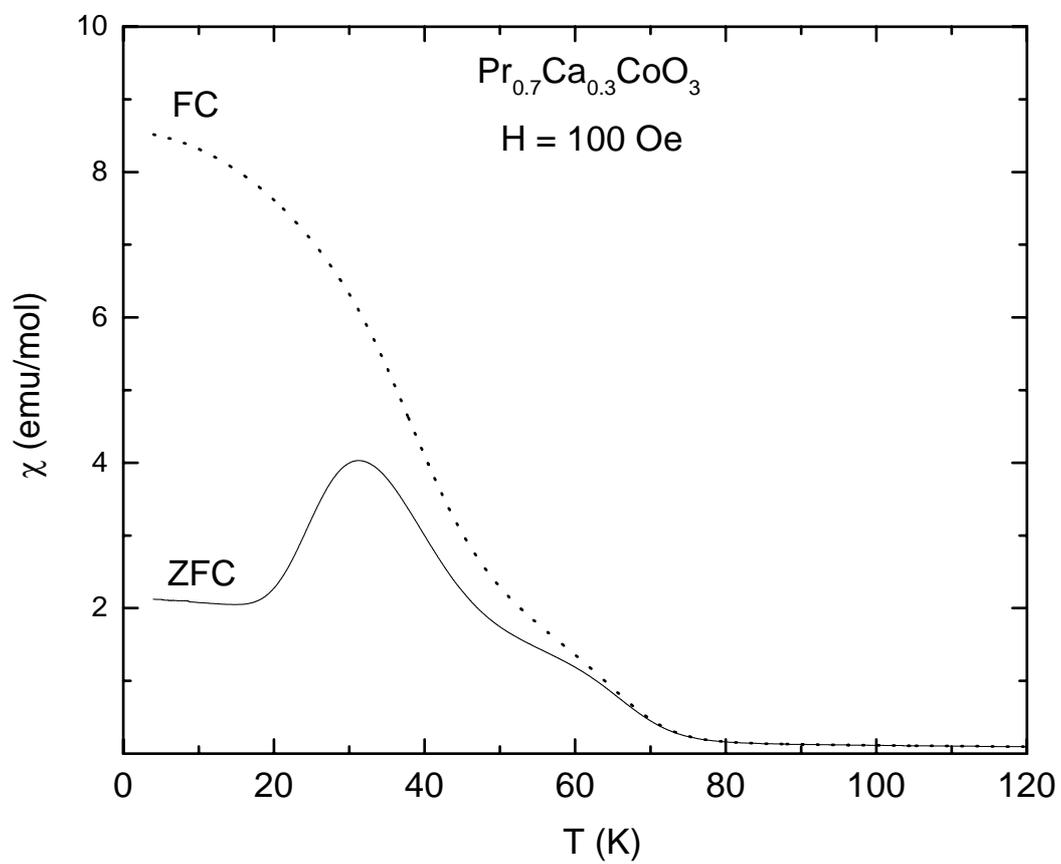

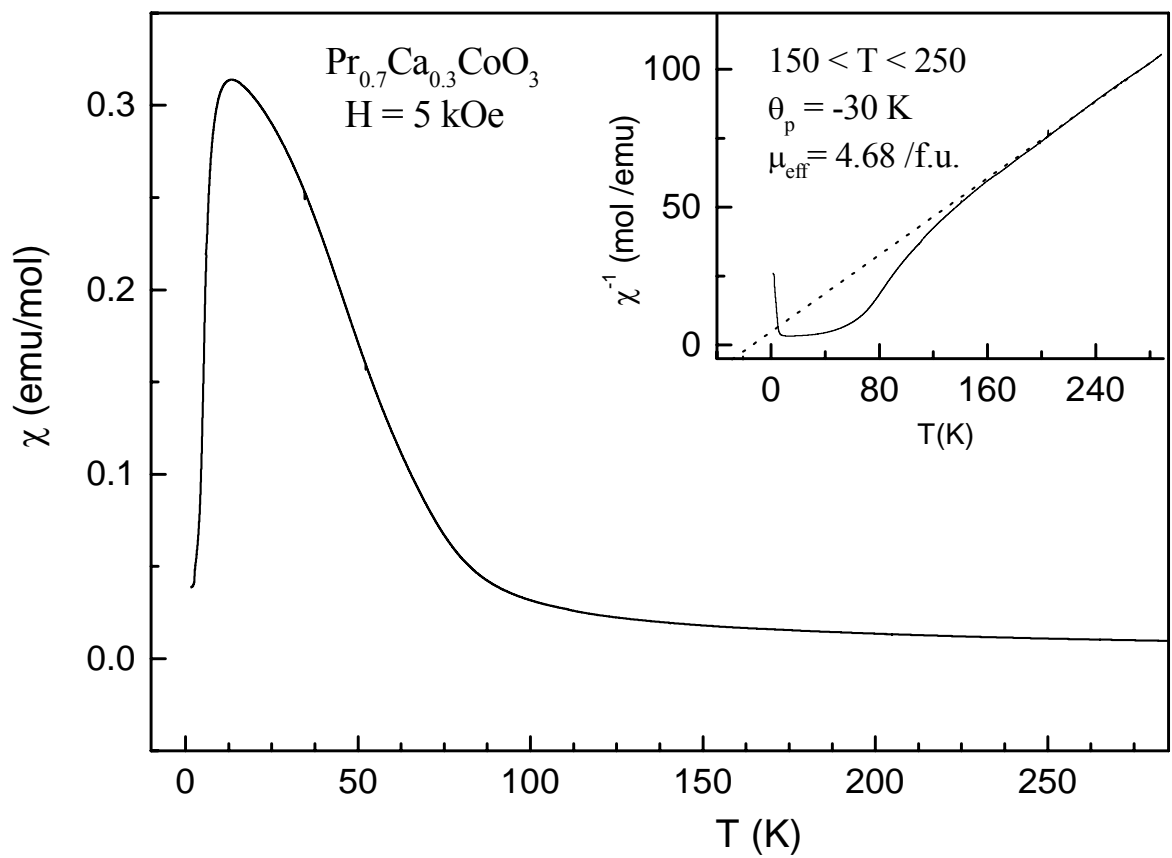

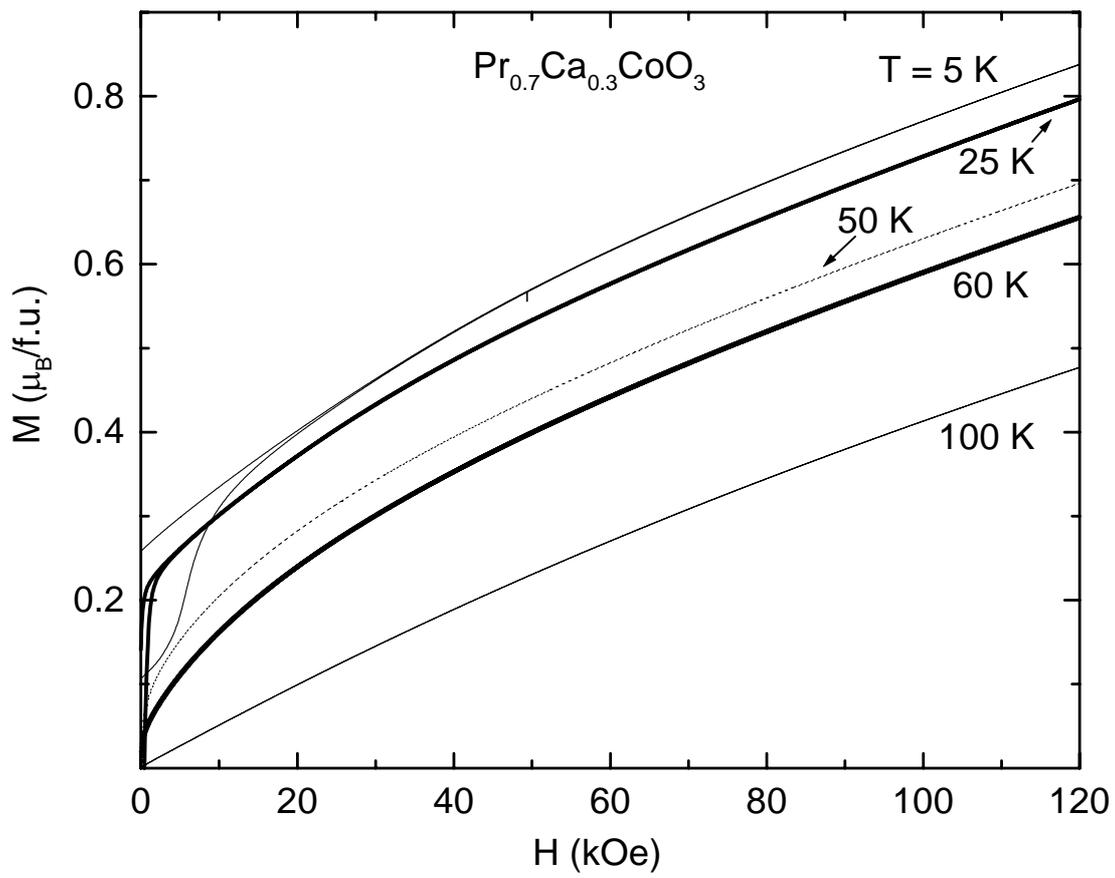

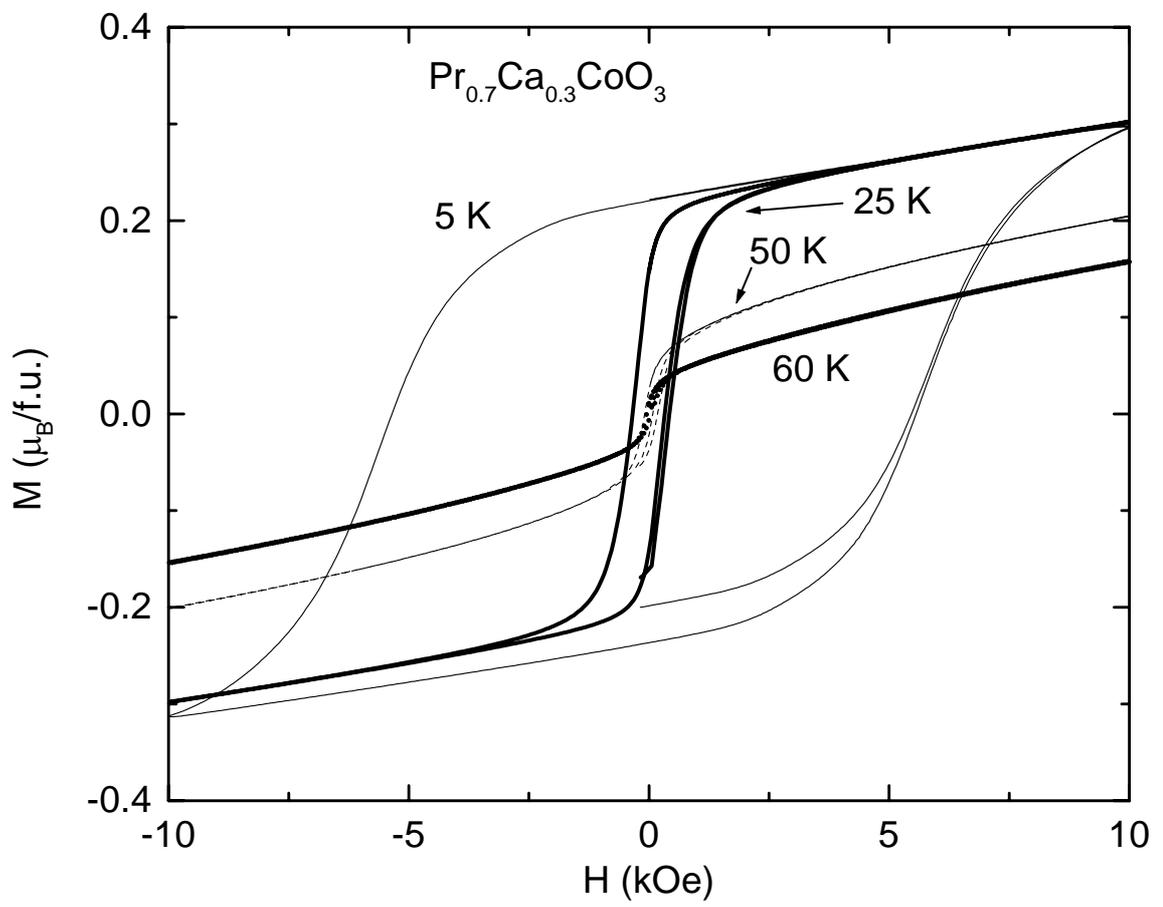

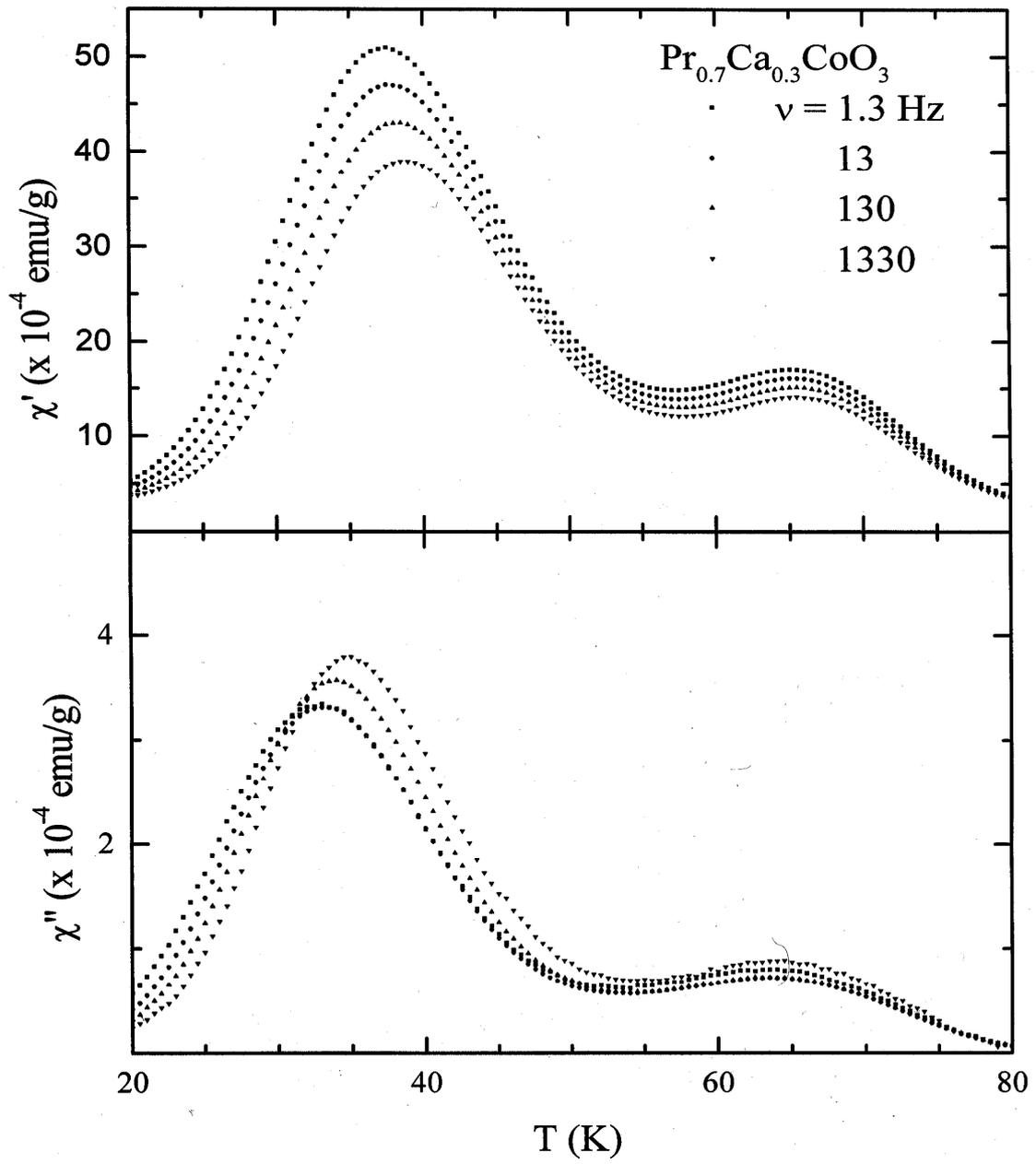

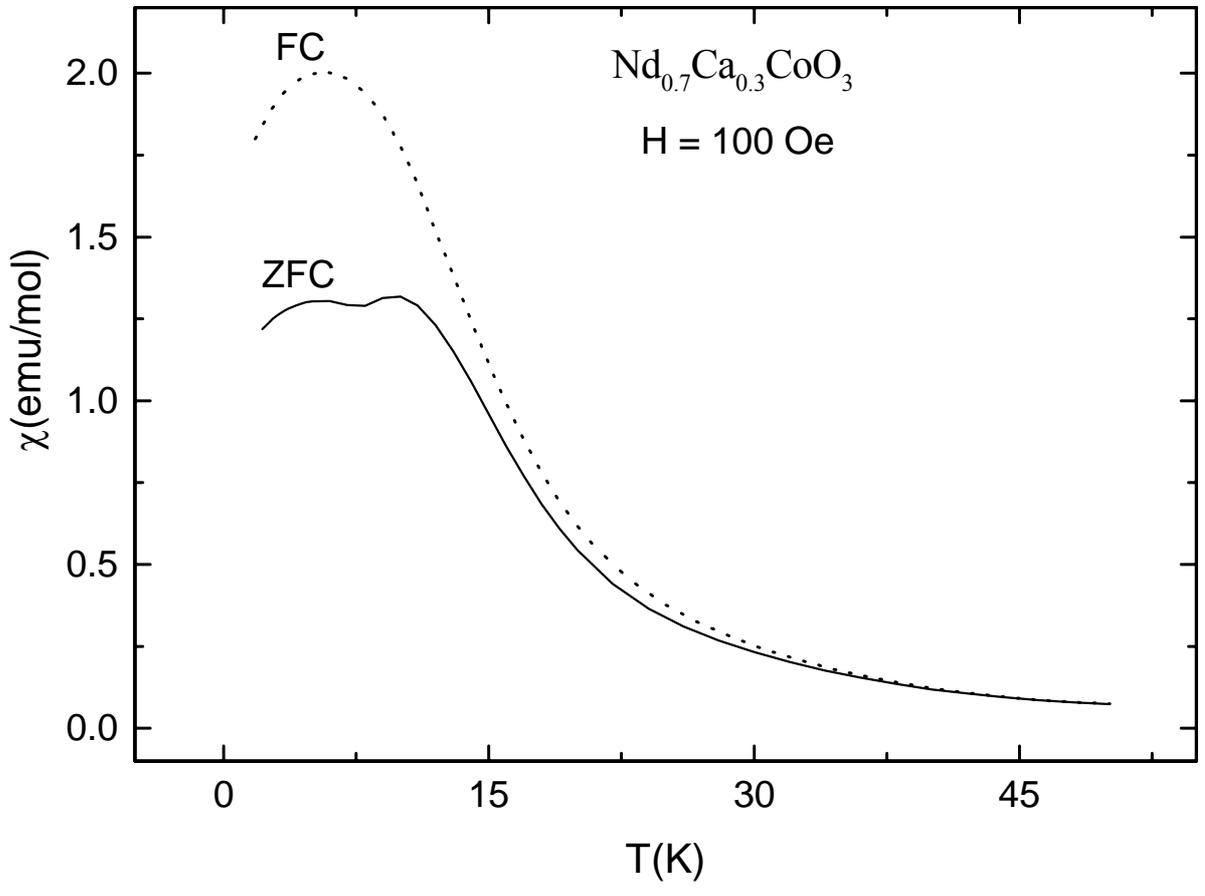

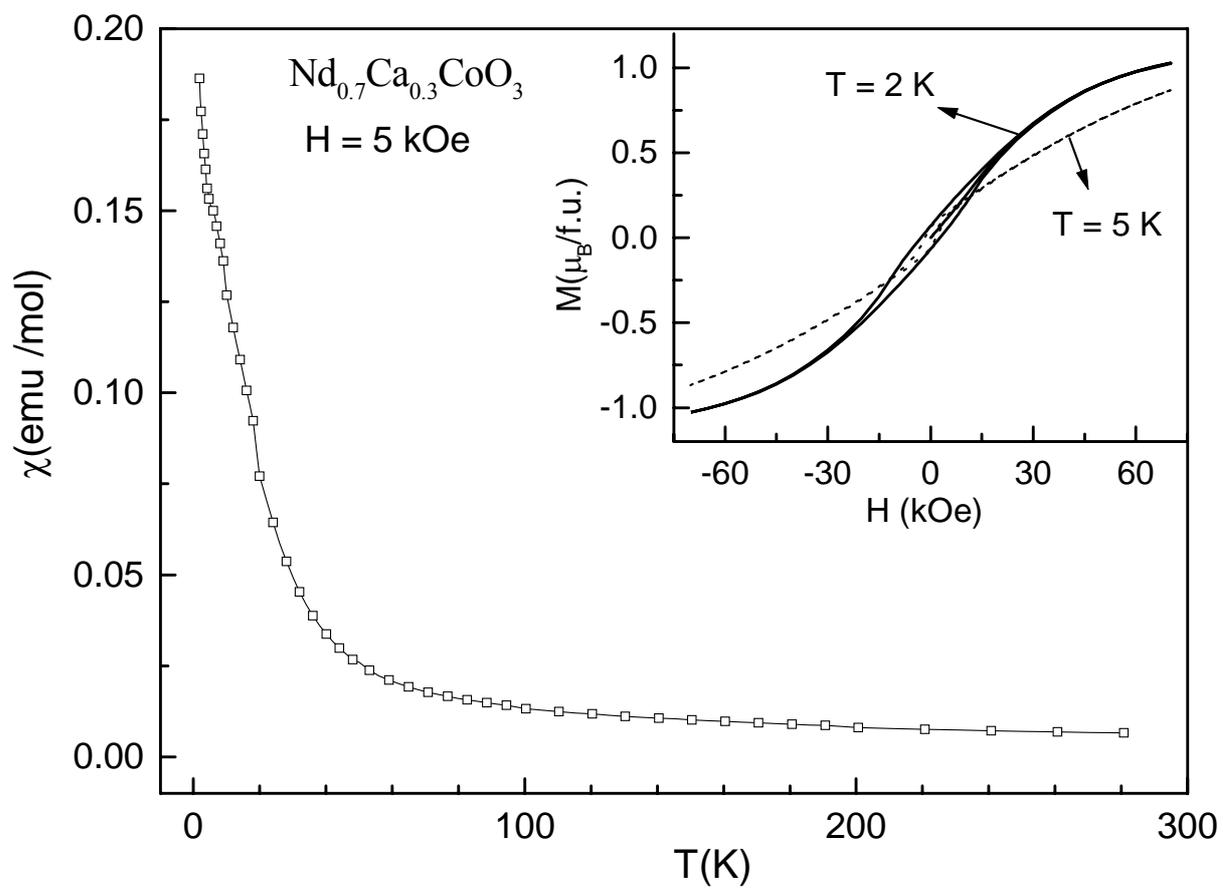